\documentclass[10pt,aps,prb,twocolumn,noshowpacs]{revtex4-1}
\usepackage{amsmath,graphicx,amsbsy,amssymb,amsmath,epsfig,latexsym,wasysym,pifont,mathrsfs,natbib}
\usepackage{float} \newfloat{widefig}{thp}{lop} \usepackage{comment}
\usepackage{array, makecell} \usepackage{verbatim} \usepackage{datetime}
\usepackage{multirow,array} \usepackage{hyperref} \usepackage{color} \usepackage{setspace}
\usepackage{subcaption}
\usepackage{graphicx}
\usepackage[export]{adjustbox}

\hypersetup{colorlinks,
  linkcolor=blue,%
  citecolor=blue,%
  urlcolor=blue}

\begin{document}

\title{\textbf{Computational studies on electrochemical performances of doped and substituted $Ti_3C_2O_2$} MXene}

\author{Mandira Das}
\email[]{mandira@iitg.ac.in}
\affiliation{Department of Physics, Indian Institute of Technology
  Guwahati, Guwahati-781039, Assam, India.}    
\author{Subhradip Ghosh}
\email{subhra@iitg.ac.in} \affiliation{Department of Physics,
  Indian Institute of Technology Guwahati, Guwahati-781039, Assam,
  India.} 
  \begin{abstract}  
Using Density functional theory (DFT) in conjunction with a solvation model we have investigated the phenomenon of eletrode- electrolyte interaction at the electrode surface and its consequences on the electrochemical properties like the charge storage and total capacitance of doped and substituted oxygen functionalised Ti$_{3}$C$_{2}$ supercapcitor electrode. We have studied nitrogen doped, nitrogen substituted and molybdenum substituted Mxenes in  acidic electrolyte H$_{2}$SO$_{4}$ solution. By considering nitrogen doping at different sites, we found that the greatest capacitance is obtained for doping at functional sites. Our results agree well with the available experiment. We also found that the enhancement in capacitances due to nitrogen doping is due to amplifications in the pseudocapcitances. We propose that the primary mechanism leading to the enhanced value of the capacitances due to nitrogen doping is surface redox activity. The performances for substituted systems, on the other hand, are degraded in comparison to the pristine ones. This suggests that better storage capacities in Ti$_3$C$_{2}$O$_{2}$ electrode can be obtained by doping only. We provide insights into the reasons behind contrasting behaviour in doped and substituted systems and suggest ways to further improve the capacitances in doped system.
\end{abstract}


\maketitle

\section{Introduction\label{intro}}
To alleviate the problem of ever increasing demand for efficient and clean energy it is imperative to design energy storage devices that meet the desired benchmarks that is a combination of high energy density, power density and long life cycle. Electrochemical capacitors or supercapacitors generated quite a bit of excitement lately due to the combination of the last two parameters found in them \cite{ren2017overview,masaki2019hierarchical,yang2020applications}. The supercapacitors can be categorised based upon the mechanism of energy storage: the electrostatic interaction giving rise to electrical double layer capacitance (EDLC) and the surface redox reaction leading to pseudocapacitance.While the EDL capacitors' electrode materials, inspite of often having large accessible surface area, can store only a few electrons per atom, pseudocapacitor electrodes can store order of magnitude higher number of electrons because of the fast and reversible redox processes. However, the accessible surfaces for the hitherto standard pseudocapacitor electrodes like RuO$_2$, MnO$_2$ and Co$_3$O$_4$ are limited \cite{ruo2,mno2,co3o4}hindering exploitation of the advantages offered. This limitation can be alleviated with the discovery of new two-dimensional (2D) nanomaterials which could maintain the high surface area while having the potential of redox reactions near the surface of the electrode material.

Accordingly, the electrochemical performances of several 2D electrodes were examined \cite{yu2015three,chen2018ultrasmall,zhang2014porous,gao2014ultrahigh,xu2018direct}.The reasonable gains obtained there were amplified once electrochemical performances of MXenes, the new members in the 2D family, were investigated. MXenes with chemical composition M$_{n+1}$X$_n$; M a transition metal, X either carbon or nitrogen,$n$ an integer, offered the necessary tunability as a plethora of new materials could have been possible by changing the chemical composition.That their surfaces are functionalized with F,O or OH \cite{functionalisation} during the process of exfoliating from MAX phases, added more possibilities towards improvement in electrochemical performances of MXene electrodes \cite{functionalisation1}. Functionalised Ti$_3$C$_2$, the first discovered MXene \cite{Ti3C2-first}, has been investigated extensively for it's electrochemical performance \cite{Ti3C2-1,Ti3C2-2,Ti3C2-3,Ti3C2-4,Ti3C2-5}. The capacitances obtained in Ti$_3$C$_2$ are much higher than those obtained from Carbon-based electrodes like Graphene \cite{Ti3C2-1}. The capacitance of few other MXene electrodes also turned out to be substantial \cite{mo2c2016,v2c2018,v2c2020}.
In order for systematic exploration of new electrode materials for supercapacitors it is important to understand the origin of the supercapacitances in the already discovered ones. Specifically, it is desirable to investigate the prevalent mechanism - EDL or Redox reaction, that leads to the superior electrochemical performance of a given electrode material. Density functional theory (DFT) \cite{dft} based first principles electronic structure calculations, in this regard, were found to have provided significant insights  \cite{kang2011,xiao2015,zhang2016,zhan2018understanding,das2022}. 

Doping and substitution with heteroatoms is a proven effective tool to improve physical properties of materials. The electrochemical performances of several 2D electrode materials like graphene \cite{dopedgraphene}, boron nitride \cite{dopedBN} and molybdenum disulfide \cite{dopedMoS2} could be significantly improved by manipulating their electron transport processes through doping.  Significant enhancement in the capcitance was observed for nitrogen doped Graphene \cite{NdopedGraphene,Ndopedgraphene1,Ndopedgraphene2,Ndopedgraphene3}. Very recently electrochemical performances of nitrogen doped Ti$_3$C$_2$ MXenes were investigated \cite{Ndopedti3c2,Ndopedti3c2-1}. In Ref ~\onlinecite{Ndopedti3c2}, improvement of capacitance by 460$\%$ was obtained at low concentration of doped nitrogen. It degraded upon increase in the nitrogen concentration. This significant increase was attributed to contributions from both EDL and redox mechanisms. A combination of DFT calculations and experimental measurements were reported in Ref ~\onlinecite{Ndopedti3c2-1}. They examined the influence of site occupancy of the dopant on the capacitance and inferred that nitrogen doping, in general increases the value of the capacitance although the gain was not as substantial as reported in Ref ~\onlinecite{Ndopedti3c2}. Their results indicated that EDL mechanism is dominant for nitrogen substitution at the carbon site while the redox mechanism prevails for doping at functional or surface sites. DFT calculations for Ti$_3$CN monolayer \cite{ti3cn}that is a 50$\%$ substitution of nitrogen at carbon lattice site suggested that nitrogen substitution may lead to significant increase in the charge storage capacity contributed mainly due to surface redox reaction. 

In the investigations on the electrochemical performances of nitrogen doped and substituted functionalised Ti$_3$C$_2$, done so far, understanding of the origin of the supercapacitative behaviour is done only qualitatively, often in an indirect way. Direct computation of various contributions to the total capacitance  providing robust ground to the qualitative understanding is still lacking. In this work, we aim at looking into this unexplored area for both doped and substituted Ti$_3$C$_2$ using DFT in conjunction with thermodynamic solvation models for accurate description of the electrochemical phenomena at the surface of the electrode in an aqueous solution. We have systematically investigated the effects on doping nitrogen at three possible sites in oxygen functionalised Ti$_3$C$_2$ (Ti$_3$C$_2$O$_2$) by computations of the different contributions to the total capacitance. Further we examined the effects of substitution at Ti and C sites by substituting 66$\%$ of Ti with Mo (Mo$_2$TiC$_2$O$_2$)and 50$\%$ C with nitrogen (Ti$_3$CNO$_2$)respectively. This systematic exploration yields a substantial understanding of the impacts of doping and substitution at different sites on the supercapacitative properties of Ti$_3$C$_2$O$_2$.  

 
\section{Methodology and Calculation Details  \label{calc_total}}
Total Capacitance($C_T$) of a Supercapacitor has contributions from the electrical part ($C_E$) and the Quantum capacitance ($C_Q$). The electrical part of the total capacitance is due to contributions from parallel combination of the EDLC ($C_{EDL}$) and the redox or pseudocapacitance ($C_{redox}$). $C_{E}$, therefore depends on the electrode-electrolyte interactions at the interface. The Quantum capacitance, on the other hand, is solely dependent on the electrode material. It originates due to imperfect screening of the electric fields by the metal electrodes and is substantial for materials at the nano-scale. Since $C_{Q}$ is in parallel with $C_{E}$ \cite{luryi1988quantum}, a non-negligible contribution from $C_{Q}$ is found to affect the overall capacitance \cite{mead,saad}.
\begin{equation}
\frac{1}{C_T} = \frac{1}{C_Q} + \frac{1}{C_E}
\label{Eqn:1}
\end{equation}

Quantum Capacitance is related to the electronic structure of the electrode material the following way, 
\begin{eqnarray}
{C_Q}^{diff} =  e^2\int_{-\infty}^{+\infty}D(\xi)F_T(\xi-U)d\xi
\label{Eqn:2} \\
{C_Q}^{int} = \frac{1}{eV}\int_{0}^{V} {C_Q}^{diff}dV
\label{Eqn:3}
\end{eqnarray}
$C_Q^{diff}$ and $C_Q^{int}$ are the differential and integrated Quantum Capacitance respectively, $D(\xi)$ is the densities of states and $F_T(\xi-U)$ is the broadening function. Integrated Quantum Capacitance gives a better representation as it provides the charge storage capacity of the electrode when charged up to a certain voltage. 

The interaction of solid electrode and liquid electrolyte leading to $C_{E}$ is simulated by the Joint Density Functional Theory (JDFT).
 JDFT is an {\it ab initio} approach that connects the quantum DFT of material with the classical DFT of liquid \cite{petrosyan2005joint}. It can accurately address the microscopic behavior of the electrostatic potential near the electrode surface. 
Here, an {\it ab initio} electrochemistry calculation is done by explicitly including the atoms making up the environment and performing thermodynamical averaging over the locations of those atoms.The free energy $A$ of an explicit quantum-mechanical system with its nuclei at fixed locations and in thermodynamic equilibrium with a liquid can be expressed as\cite{petrosyan2007joint},
\begin{multline}
A = \min_{n(r),\big\{N_{\alpha}(r)\big\}} \Bigg\{G[n(r),{N_{\alpha}(r)},V(r)] - \\
 \int d^3rV(r)n(r)\Bigg\}
 \label{Eqn:4}
\end{multline} 
 $G[n(r),\big\{N_{\alpha}(r)\big\},V(r)]$ is an universal functional of $n(r)$, the electron density of the explicit system, $\big\{N_{\alpha}(r)\big\}$, the densities of the nuclei of the various atomic species in the environment , and $V(r)$, the electrostatic potential due to the nuclei of the explicit system. $G[n(r),\big\{N_{\alpha}(r)\big\},V(r)]$ depends only on the nature of the environment and that its dependence on the explicit system is only through the electrostatic potential of the nuclei included in $V(r)$ and $n(r)$, the electron density of the explicit system. The functional $G[n(r),\big\{N_{\alpha}(r)\big\},V(r)]$ can be separated into three parts\cite{petrosyan2007joint} 
\begin{multline}
G[n(r),\big\{N_{\alpha}(r)\big\},V(r)] \equiv A_{KS}[n(r)] + \Omega_{lq}[\big\{N_{\alpha}(r)\big\}] + \\
\Delta A[n(r),\big\{N_{\alpha}(r)\big\},V(r)]
\end{multline}
$A_{KS}[n(r)]$ and $\Omega_{lq}[\big\{N_{\alpha}\big\}]$ are the standard universal Kohn-Sham electron-density functional of the explicit solute system in isolation(including its nuclei and their interaction with its electrons) and the 'classical' density functional for the liquid solvent environment in isolation, respectively. $\Delta A[n(r),\big\{N_{\alpha}(r)\big\},V(r)]$ depicts the coupling between the solvent and the solute.The solvation effect and the electrolyte response are then approximated by implicit solvation models\cite{letchworth2012joint}. The free energy  in this approximation is given as
\begin{multline}
\tilde{A} = \min_{n(r)} (A_{KS}[n(r),\big\{Z_I,R_I\big\}] + \Delta \tilde{A}[n(r),\big\{Z_I, R_I\big\}])
\label{Eqn:6}
\end{multline}
The effects of the liquid environment appears in the new term,
\begin{multline}
\Delta \tilde{A}[n(r),\big\{Z_I,R_I\big\}] \equiv \min_{N_{\alpha}(r)}(\Omega_{lq}[N_{\alpha}(r)] + \\
\Delta A[n(r),N_{alpha}(r),\big\{Z_I,R_I\big\}])
\label{Eqn:7}
\end{multline}
$Z_I$ and $R_I$ are the charges and positions of the surface nuclei. Minimization of equation (\ref{Eqn:6}) leaves a functional in terms of only the properties of the explicit system and incorporates all of the solvent effects implicitly. There are many advanced solvation models for JDFT which include non-local effects\cite{kornyshev1998nonlocal}, nonlinear dielectric response\cite{gunceler2013importance}, spherically averaged liquid susceptibility ansatz (SaLSA)\cite{sundararaman2015spicing}, and charge-asymmetric nonlocally determined local-electric (CANDLE)\cite{sundararaman2015charge} solvation model.

In this work, we have considered a non-ideal Faradaic process as implemented by Gogotsi $et$ $al.$ \cite{zhan2018understanding} to compute the capacitances due to EDL and surface redox reactions for $Ti_3C_2O_2$ in $H_2SO_4$ electrolyte. We have generalised it for our systems. 
The redox reaction for N-doped $Ti_3C_2O_2$ (N-$Ti_3C_2O_2$) can be written as ,
\begin{multline}
    {N-Ti_3C_2O^{q}_2} + 2H^{+}(aq) \\ 
    \longrightarrow 
    {N-Ti_3C_2O_{2-2x}(OH)^{q^{'}}_{2-2x}} + 2(1-x)H^{+}(aq)
    \label{Eqn:8}
\end{multline}
where $x$ is surface H coverage (between 0 and 1) and $q$ is the net charge on the electrode.The applied voltage controls both the $q$ and $x$ during the charging process. The free energy of the charged electrode system with coverage ($x$) and applied electrode potential$(\phi)$ is given by
\begin{multline}
   G(x,\phi) = E(x) + xE_{ZPE} + Q(V(x,\phi))\phi \\
    + E_{EDL}(V(x,\phi)) + (1-x)\mu_{H^{+}}
   \label{Eqn:9}
\end{multline}
$E(x)$ is the total energy of the solvated electrode with H coverage of $x$ in zero surface charge. $E_{ZPE}$
is zero point energy difference of the electrode between no H and full H coverage.The term, $Q\phi$ , is the electrical work to move the charge $Q$ (net charge on electrode) form zero potential (in the bulk electrolyte) to the electrode with the potential $\phi$.$E_{EDL}$ is the energy of the induced EDL by the electrode charge $Q$.The last term is the chemical potential of the solvated proton in the electrolyte. $V(x,\phi)$ is the relative potential with respect to PZC (Point of zero charge) at coverage $x$ and electrode potential $\phi$,
\begin{equation}
    V(x,\phi) = \phi - \phi_{PZC}(x)
    \label{Eqn:10}
\end{equation}
Once $V(x,\phi)$ is known , the charge $Q$ and $E_{EDL}$ can be obtained by ,
\begin{equation}
    Q(x,\phi) = \int_{V=0}^{V=\phi-\phi_{PZC}(x)} C_{EDL}dV
    \label{Eqn:11}
\end{equation}
\begin{equation}
    E_{EDL}(x,\phi) = \int_{V=0}^{V=\phi-\phi_{PZC}(x)} Q(x,\phi) dV
    \label{Eqn:12}
\end{equation}
The final term $\mu_{H^{+}}$ (proton's chemical potential) is given by,
\begin{equation}
    \mu_{H^{+}} = \frac{1}{2} G[H_2] + e\Phi_{SHE} - 0.059 \times pH
    \label{Eqn:13}
\end{equation}
where $G[H_2]$ is 
\begin{equation}
    G[H_2] = E[H_2] + ZPE[H_2] + \frac{7}{2}k_B T - TS_{H_2}
    \label{Eqn:14}
\end{equation}
All the physical quantites in Equation (\ref{Eqn:14}) can be obtained by DFT calculations and standard thermodynamic database. To this end, we have used the JDFT method with an implicit solvation method as implemented in the simulation package JDFTx\cite{sundararaman2017jdftx} to obtain the electronic structure and the potential at the PZC of each H coverage for calculation of the Free energy function. The implicit electrolyte is described by the charge-asymmetric non-locally determined local-electric (CANDLE) model. $\Phi_{SHE}$ is the computational standard hydrogen electrode. It has been determined to be 4.66 V from PZC calibration of the CANDLE solvation in JDFTx\cite{sundararaman2015charge}.$E(x)$ and PZC is determined form the JDFTx simulation for different $x$ coverages and configurations. We fit the data to obtain  relationships of $E(x)$ and PZC with $x$ and feed them to  Equations (\ref{Eqn:2}) and (\ref{Eqn:3}) to calculate relative free energy with H-coverage at any given electrode potential. Once the relative free energy is obtained,H-coverage $x$ and net surface charge $Q$ are calculated by taking ensemble averages. As there are different H coverage configurations for any given value of coverage , degenerate statistical mechanics is used to calculate the ensemble averages.The charge associated with the H-coverage leads to the redox capacitance and the one associated with surface charge $Q$ yields the EDL capacitance. The slope of the charge with electrode potential gives total electric capacitance ($C_E$) at any given electrode potential.
The Generalised Gradient Approximation with the Perdew-Burke-Enzerhof functional (GGA-PBE)\cite{perdew1996generalized} was considered as the approximate electron exchange-correlation for the DFT calculations. Ultrasoft pseudopotentials\cite{garrity2014pseudopotentials} are used to describe the ion-electron interaction. Kinetic energy cut-off of 20 Hartree and 30 Hartree were used for structure optimization and single point energy calculation at the optimized geometry with convergence criteria of $10^{-6}$ Hartree. For all calculations,  $3 \times 3 \times 1$ supercells of the unit cells were considered.

\section{Results And Discussions}
\subsection{Structural models and properties}
\begin{figure}[h]
 \center
   \begin{subfigure}{0.5\textwidth}
     \centering
      \includegraphics[scale=0.30]{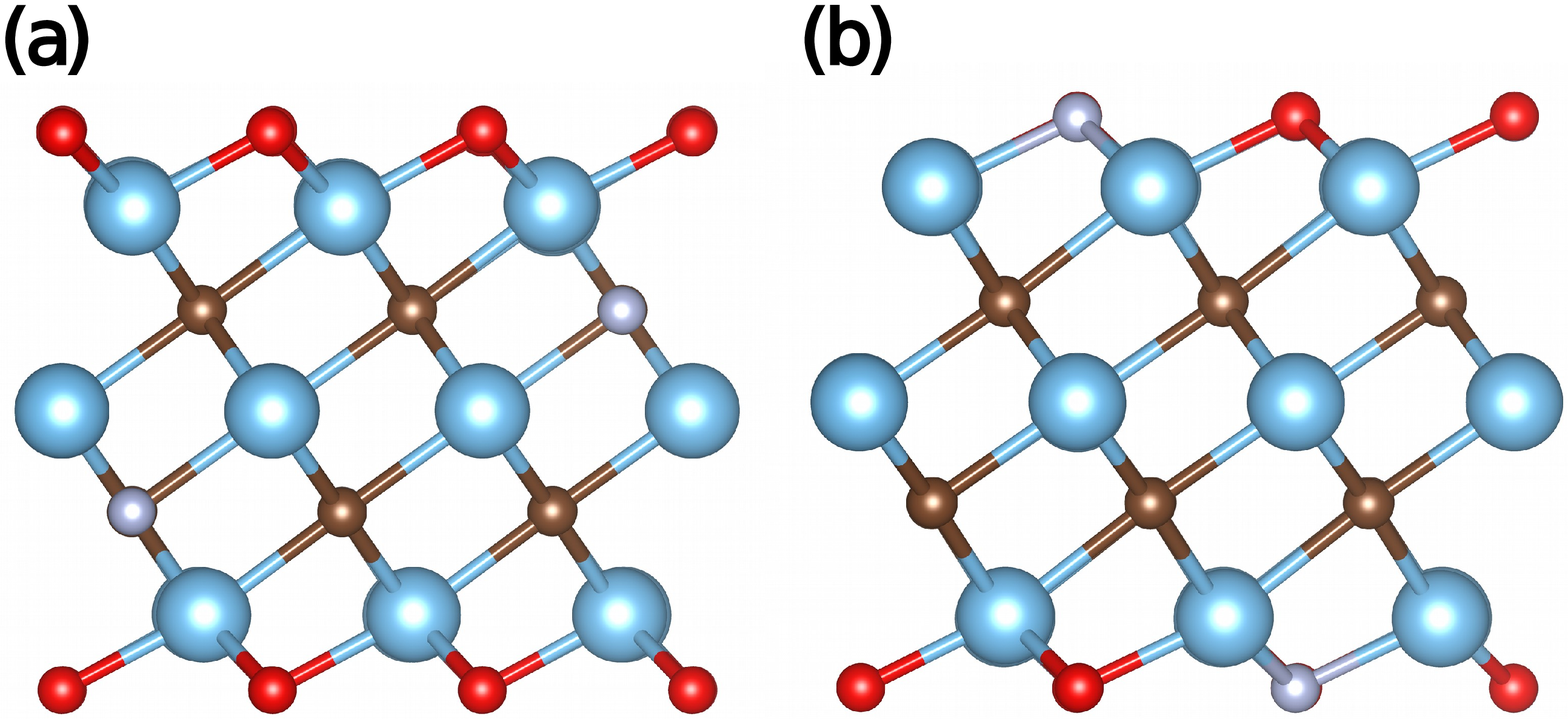}
   \end{subfigure}
    \begin{subfigure}{0.5\textwidth}
      \centering
       \includegraphics[scale=0.30]{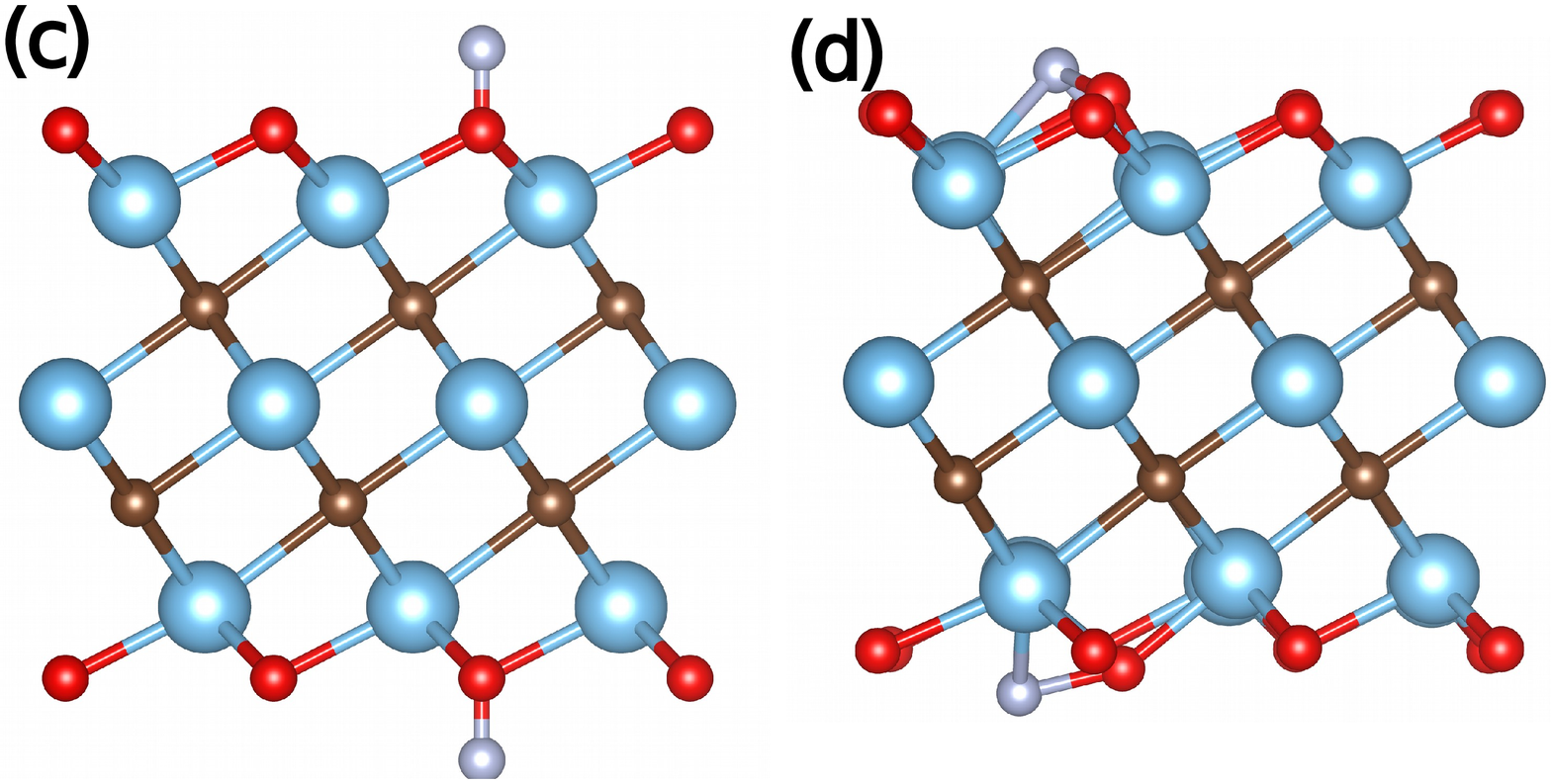}
  \end{subfigure}
   \caption{Structural models of Nitrogen doped $Ti_3C_2T_x$: (a)LS model , (b)FS model , and (c)SS model, where N is located above the O atom , (d)SS model, where N is located between O and Ti atoms.Blue , brown, red and sky-blue balls represent Ti, C, O and N atom respectively. }
   \label{Fig:1}
 \end{figure}

Ti$_3$CNO$_2$ and Mo$_2$TiC$_2$O$_2$ are modelled by replacing 50$\%$ of carbon layer (of Ti$_3$C$_2$O$_2$) by nitrogen and by replacing the outer Ti layers (of Ti$_3$C$_2$ O$_2$)by Mo, respectively. Modelling the nitrogen doped Ti$_3$C$_2$O$_2$ is, however, different. It is because the nitrogen atom can occupy three possible positions when doped into Ti$_3$C$_2$O$_2$ MXene. Accordingly, three structural models can be constructed:  Lattice-Site(LS) model, Functional-Site(FS) model, and Surface-Site(SS) model. The doped nitrogen replace carbon and Oxygen atoms in LS and FS models, respectively. In the SS model, they do not replace any atoms, but position themselves on the surface of the MXene. They can occupy either the positions on top of the functional atoms or the positions between the functional  and  Ti atom in the outer layers. The structural models are shown in Figure \ref{Fig:1}.
 In this work we have considered 10\% nitrogen doping in Ti$_3$C$_2$O$_2$. Total energy calculations are done to obtain optimised structures for the FS and LS doping models. The SS model was not considered as calculations in Reference ~\onlinecite{Ndopedti3c2-1} showed it to be dynamically unstable.
The structural parameters  of pristine and doped Ti$_3$C$_2$O$_2$ are given in Table-\ref{tab:1}. The results are in excellent agreement with the experimental ones \cite{Ndopedti3c2-1}.
\begin{table}[h]
    \centering
    \caption{Optimised lattice parameters and bond lengths of pristine and nitrogen doped $Ti_3C_2T_x$ in {\AA}}
    \begin{tabular}{c@{\hspace{0.4cm}} c@{\hspace{0.4cm}} cccc@{\hspace{0.8cm}}} 
    \hline\hline
    \vspace{-0.33cm}
    \\ System & Lattice parameter  &  \multicolumn{3}{c}{Bond lengths}\\
        &     &    Ti-C & Ti-O & Ti-N  \\
    \hline
    Un-doped       &  3.03  &   2.19  &  1.97 & -    \\
    Lattice-Site  &  3.01  &   2.12  &  1.99 & 2.20   \\
    Functional-Site &  3.04 & 2.26  & 1.96 & 1.91    \\
    \hline
    \end{tabular}
    \label{tab:1}
\end{table}
We find that the lattice  contracts upon doping at the lattice site and expands upon doping at the functional site. This trend in the lattice parameters is consistent with the size of the atoms. The atomic radius of Oxygen, Carbon and Nitrogen are 60pm,70pm and 65pm respectively. The changes in the lattice parameters upon doping are consistent with this. The length of the Ti-C bond expands(contracts) in the FS(LS) model, in comparison to that in the pristine system. A reverse trend is shown in the case of Ti-O bond lengths, although the changes in this case are almost negligible. The Ti-N  bond length changes substantially between the two models. This is expected since different atoms are replaced in different models. It suggests that the charge sharing between Ti and N will be more in case of FS model than the LS one. Overall, the trends in the lattice parameters follow the trends in the Ti-C bond lengths. 
\subsection{Electronic Structure and Quantum Capacitance}
\begin{figure}[h]
    \vspace{-0.2cm}
    \includegraphics[width=0.5\textwidth,center]{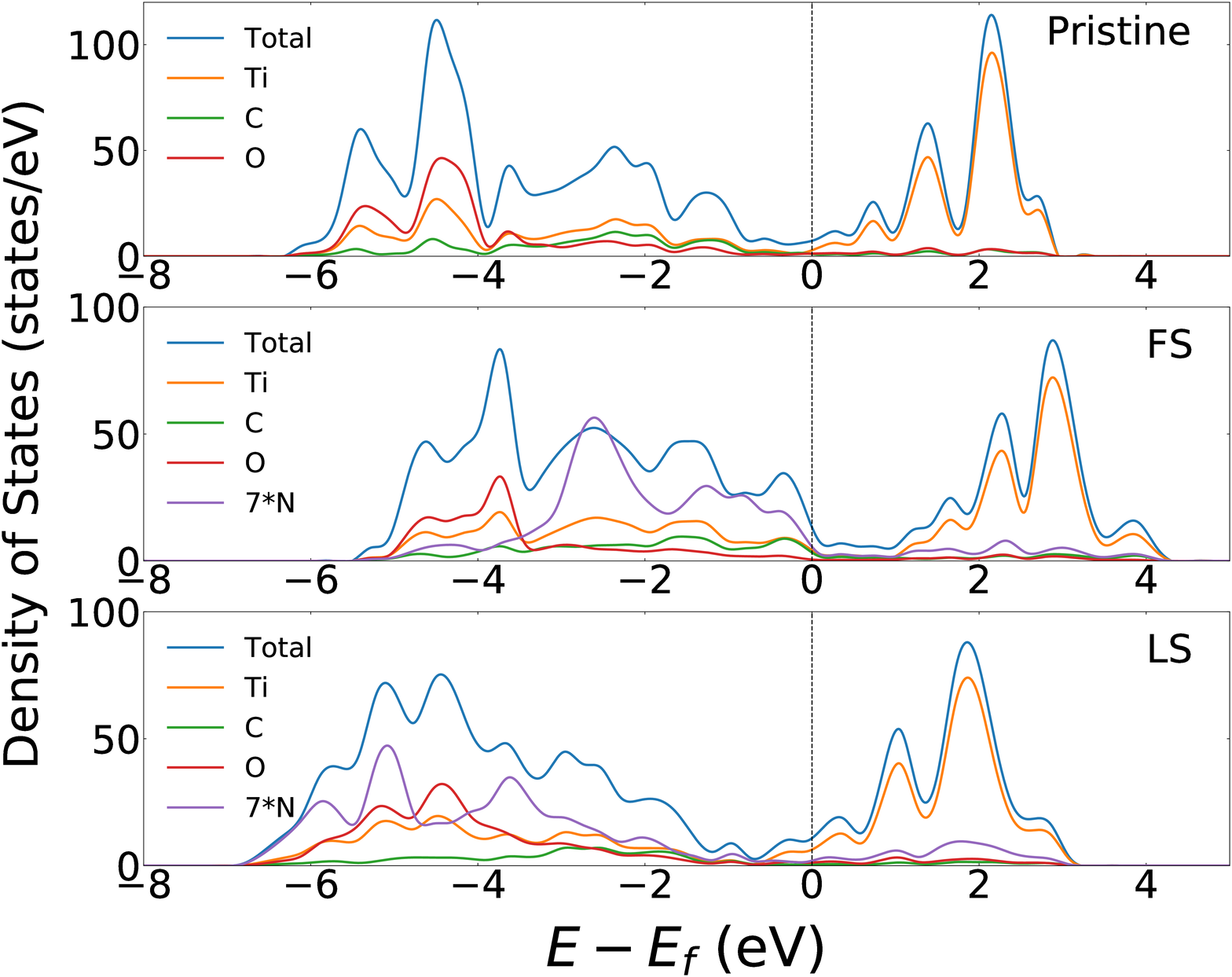}
    \caption{Electronic Structure of pristine (upper panel),FS doped (middle panel) and LS doped $Ti_3C_2O_2$(lower panel)}
    \label{Fig:2}
\end{figure}

The quantum capacitance of an electrode material depends on its electronic structure near the Fermi level. 
In Figure \ref{Fig:2} we present the electronic structure of pristine and nitrogen doped (both FS and LS) $Ti_3C_2O_2$. The nitrogen atom has one more valence electron than carbon and one less electron than the oxygen. Therefore, when a nitrogen atom replaces a carbon(oxygen) atom,  total number of electrons in the system increases (reduces).As a consequence, the electronic bands around the Fermi-levels are shifted. We find that in case of doping at the functional site (middle pnel of Figure \ref{Fig:2}), the electronic band shifts towards higher energies. This happens due to the reduction in the number of electrons in the system. In case of doping at the lattice site, the shift of the electronic bands is towards lower energies(lower panel of Figure \ref{Fig:2}). This is due to increase in total number of electrons in the system.
Close inspection of the densities of states reveals the following features: (a) the electronic structure near the Fermi level is most substantially affected in case of doping at the functional site. This is mostly due to the nitrogen $p$ states occupying higher energy levels in the valence band (b) in case of doping at the lattice sites, the nitrogen $p$ states are deep within the valence band unable to make an impact on the features near the Fermi level. The features near the Fermi level, in this case, are mostly due to the Ti $d$ states (c) the unoccupied part of the spectra are overwhelmingly due to Ti $d$-bands which implies serious consequences on the charge transfer. From these we can predict that it is more likely that the largest contribution to $C_{Q}$ will come from the functional site doped system.  

\begin{figure}[h]
    \includegraphics[width=0.5\textwidth,center]{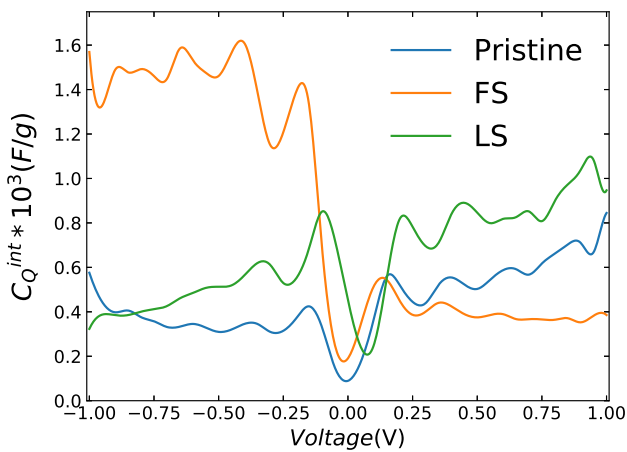}
    \caption{Integrated Quantum Capacitance of pristine and doped $Ti_3C_2O_2$}
    \label{Fig:3}
\end{figure}

In Figure \ref{Fig:3}, we present the variation of  Integrated Quantum capacitance($C_Q^{int}$) with absolute voltage. The voltage range is set between $\pm$1 V. This range is chosen as because at room temperature, the electrochemical stability window of the electrolyte solvent is about 1.25 eV \cite{stability}. In the negative part of the voltage window, maximum $C_Q^{int}$ of FS doped Ti$_3$C$_2$O$_2$ is four times larger than that of the pristine MXene. This comparative behaviour correlates to the changes in the electronic structure near the Fermi level. 
As expected from the features in the electronic structure, $C_Q^{int}$ of LS doped system, though higher than the pristine system throughout the voltage window, is not as amplified as the FS doped system.

\begin{figure}[h]
    \includegraphics[width=0.5\textwidth,center]{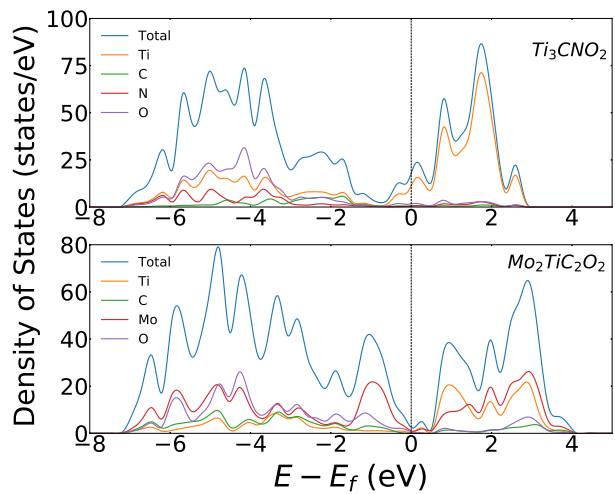}
      \caption{Electronic Structure of $Ti_3CNO_2$(Top) and $Mo_2TiC_2O_2$(Bottom)}
    \label{Fig:4}
\end{figure}

Figure \ref{Fig:4} shows electronic structures of $Ti_3CNO_2$ and $Mo_2TiC_2O_2$. A comparison between LS doped Ti$_3$C$_2$O$_2$ and Ti$_3$CNO$_2$ can be instructive. One of the prominent features is that with introduction of more nitrogen replacing carbon, the bands move towards lower energy, in general. A consistent trend emerges as one makes a comparison between pristine, LS doped and carbon substituted Ti$_3$C$_2$O$_2$. In Ti$_3$CNO$_2$, states near Fermi level are populated due to more contributions from nitrogen. The shoulder at around 3 eV in LS doped Ti$_3$C$_2$O$_2$  transforms to a peak in Ti$_3$CNO$_2$. The small peak closest to the Fermi level, in the unoccupied part in LS doped system transforms into a prominent one with nitrogen substitution. Upon substitution of Mo at Ti positions, however, the electronic structure changes considerably. The densities of states in Mo$_2$TiC$_2$O$_2$  is dominated by contributions from Mo and O near the Fermi level in the occupied part of the spectrum. In the unoccupied part, states near the Fermi level have dominant contributions from Ti and Mo. The bands are more extended in Mo$_{2}$TiC$_{2}$O$_{2}$ in comparison to Ti$_3$C$_2$O$_2$ as Mo being from the $4d$ series have wider bands. These features reflect in variations in $C_Q^{int}$.  For $Ti_3CNO_2$, the higher density of states in the conduction band around the Fermi-level leads to significantly larger $C_Q^{int}$  in the positive side of the voltage window in comparison with that in the negative side. As expected from the features of the densities of states, $Mo_2TiC_2O_2$ have the lowest integrated quantum capacitance throughout the voltage window. 

\begin{figure}[h]
    \includegraphics[width=0.5\textwidth,center]{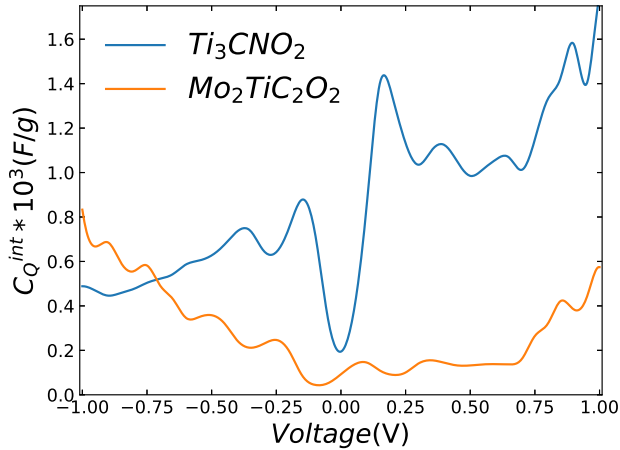}
    \caption{Integrated Quantum Capacitance of $Ti_3CNO_2$ and $Mo_2TiC_2O_2$}
    \label{Fig:5}
\end{figure}

Equation (\ref{Eqn:1}) implies that larger value of quantum capacitance will lead to higher value of the total capacitance where the electrical part of the total capacitance will be the sole contributor. If both electrical and quantum parts of the capacitance are comparable, the total capacitance will be lesser than the least. Our results suggest that in comparison to others, FS doped Ti$_3$C$_2$O$_2$ (Ti$_3$CNO$_2$) may lead to higher total capacitances when operated in the negative(positive) voltage windows. A final call on this will be possible only when the electrical part of the capacitances are calculated.In the next sub-sections we explore this point.

\subsection{Electrical capacitance}
As discussed in Section \ref{calc_total}, we have used the model of MXene electrode in contact with an acidic electrolyte H$_2$SO$_{4}$ for calculation of the EDLC and Pseudocapacitances. We specifically looked at the effects of doping and substitution at different sites on the evolution of the electron and proton transfers between the electrolyte and the electrode, giving rise to simultaneous contributions of EDL and redox capacitances. The literature on nitrogen doped functionalised Ti$_3$C$_2$ has proposed EDL as a primary mechanism behind elevated value of the total capacitance due to nitrogen doping \cite{Ndopedti3c2,Ndopedti3c2-1}. On the other hand, analysis based solely on the electronic structures of  functionalisedTi$_3$CN monolayers inferred that the primary mechanism for O functionalised system is the surface redox reaction while for F and (OH) functionalisation it is EDL \cite{ti3cn}. One of our motivations, therefore, is to investigate this aspect.
\subsubsection{Nitrogen Doped Ti$_3$C$_2$O$_2$}
\begin{figure*}
    \includegraphics[width=1.2\textwidth,center]{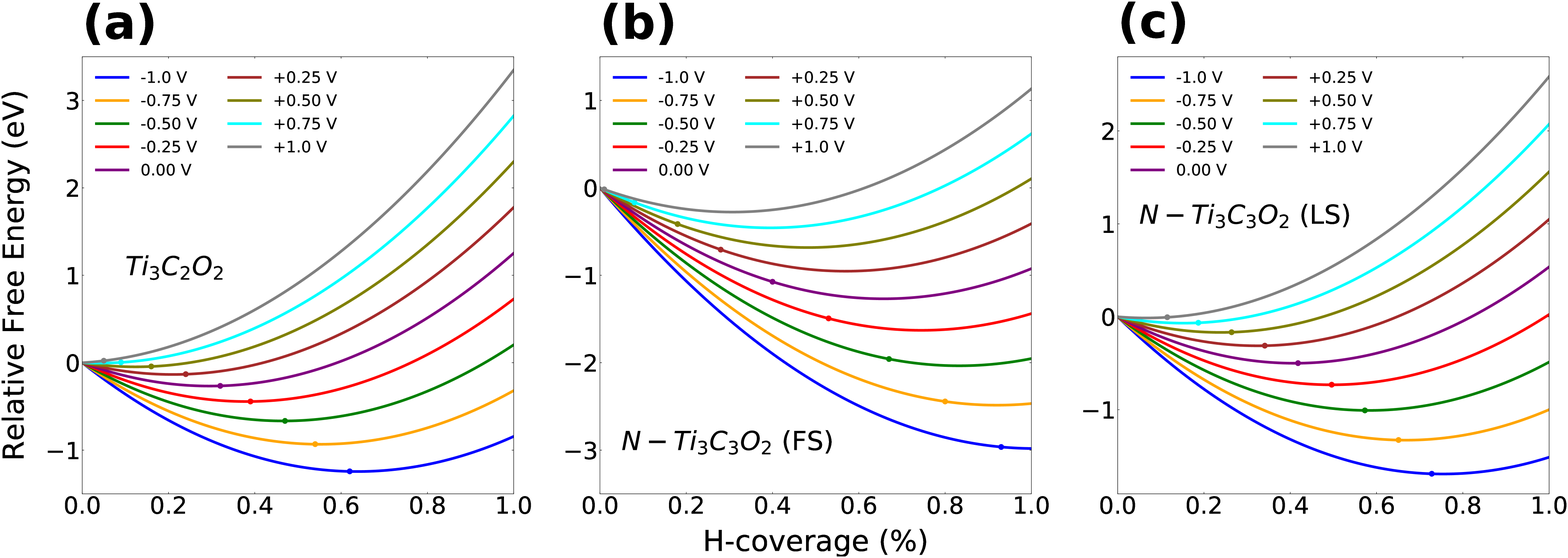}
    \vspace{0.1cm}
    \includegraphics[width=1.2\textwidth,center]{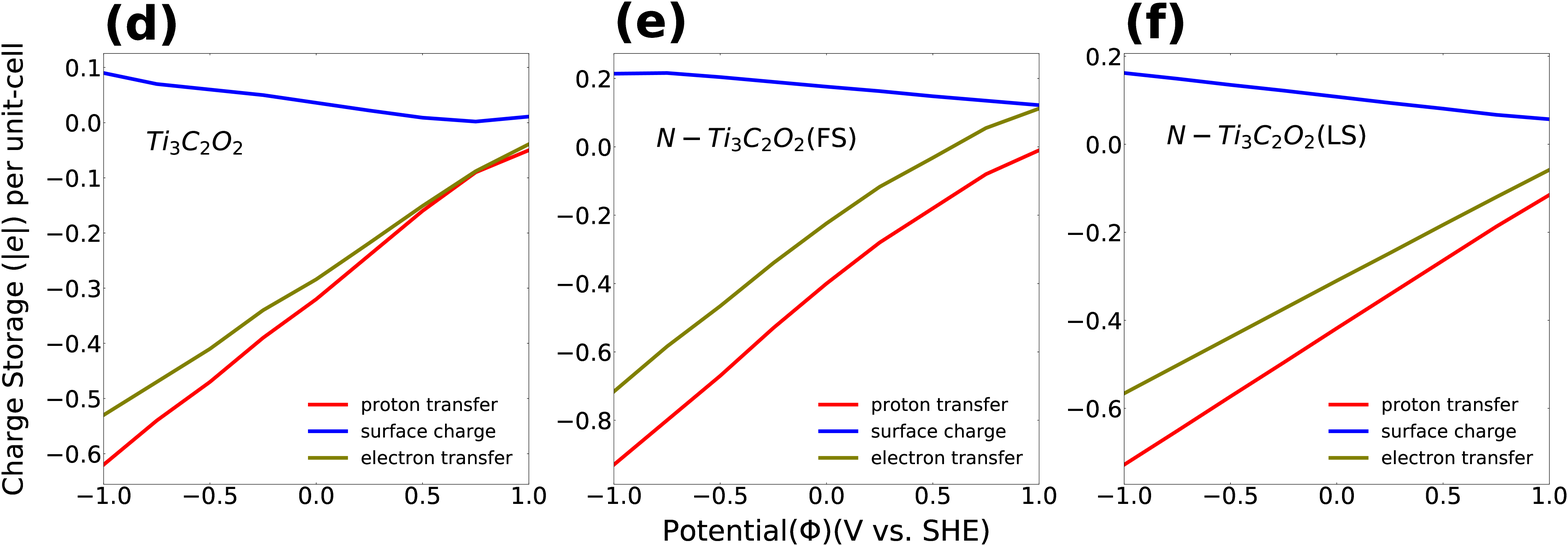} 
    \vspace{0.1cm}
    \includegraphics[width=1.2\textwidth,center]{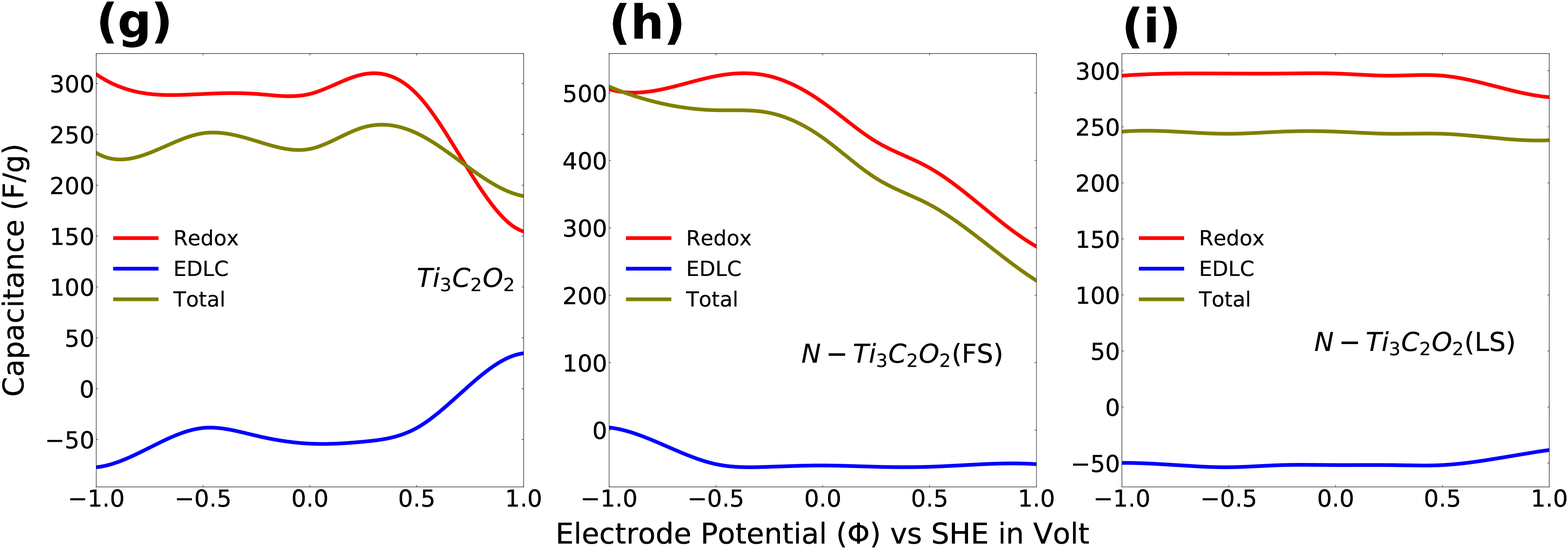}
   \caption{Variations in the Relative Free Energy with H-coverage at different fixed potentials for pristine(a), nitrogen doped FS (b) nitrogen doped LS (c ) $Ti_3C_2O_2$. The charge storage and electrical capacitances as a function of electrode potential are shown for pristine (d,g), nitrogen doped FS (e,h) nitrogen doped LS (f,i) systems. The negative values of the EDL capacitances imply that the EDL mechanism is acting against the surface redox mechanism.}
    \label{Fig:6}
\end{figure*}

Acoording to Equation (\ref{Eqn:8}), the surface redox behaviour is controlled by adsorption of proton on the electrode surface and subsequent electron transfer to the electrode at a given electrode potential. To determine the extent of the surface redox phenomenon we calculated $G(x,\phi)$ (Equation \ref{Eqn:9}) for each H-coverage, $x$, at different fixed electrode potentials ($\Phi$); $\Phi$ varying from -1 V to +1 V with respect to SHE.  
In Figure-\ref{Fig:6} (a)-(c), we show $G(x,\phi)$ as a function of $x$ for pristine, nitrogen doped FS and LS $Ti_3C_2O_2$ , respectively. For all systems, and for all $\Phi$, $G(x, \Phi)$ is  parabolic in nature. From the minima of $G(x,\Phi)$, we find that the ensemble averaged coverage $x_{avg}= 0.63  , 0.93. 0.73$ for pristine, functional-site and lattice-site doped $Ti_3C_2O_2$, respectively. The result for pristine compound agrees very well with the existing one \cite{zhan2018understanding}. The results imply that the electrode surfaces are not fully protonated even at -1 V for all systems.However, nitrogen doping enhances protonation with the FS system having more than 90\% covered, the maximum among the three systems. Another common feature in all systems is that the $x_{avg}$ continuously decreases with increasing $\Phi$ leading to $x_{avg}=0$ at $\Phi=1$ V.
The electrochemical behaviour of a capacitor in acidic electrolyte like H$_2$SO$_4$ can be understood in terms of the net charge on the surface that is responsible for EDL, and the proton transfer from the electrolyte ions to the electrode that is responsible for the redox reaction. The net number of electron transfer or the total charge storage in the electrode is decided by the cumulative effect of these two.
The electron transfer, surface charge, and proton transfer numbers as a function of $\Phi$ are shown in the Figure-\ref{Fig:6} (d)-(f) for pristine, nitrogen doped FS and LS systems. The proton transfer number is given by $x_{avg}$ while the surface charge is the ensemble average of $Q$ in Equation (\ref{Eqn:11}). The electron transfer happening across the charge layer accumulated over the electrode surface is obtained by adding these two. We find that for all three systems accumulated surface charge prohibits the proton transfer to the surface. For the LS doped system, the variations in surface charge, proton transfer and consequently the electron transfer numbers are linear with $\Phi$. The surface, in all three cases, are slightly positively charged. This is due to the fact that part of the positive charge from proton transfer becomes the net surface charge. In case of the pristine and the FS doped systems, the variations are piecewise linear with changes in slopes at intermediate voltages, the slopes being larger for FS doped system. The pseudo-capacitance, the EDLC and the total electrical capacitance are obtained from the slopes of proton transfer number, surface charge, and electron transfer number curves (Fig~\ref{Fig:6}(d-f)) respectively.Figure \ref{Fig:6} (g)-(i) shows variations in the capacitances for the three systems under consideration.
\begin{figure*}[ht]
    \includegraphics[width=0.95\textwidth,center]{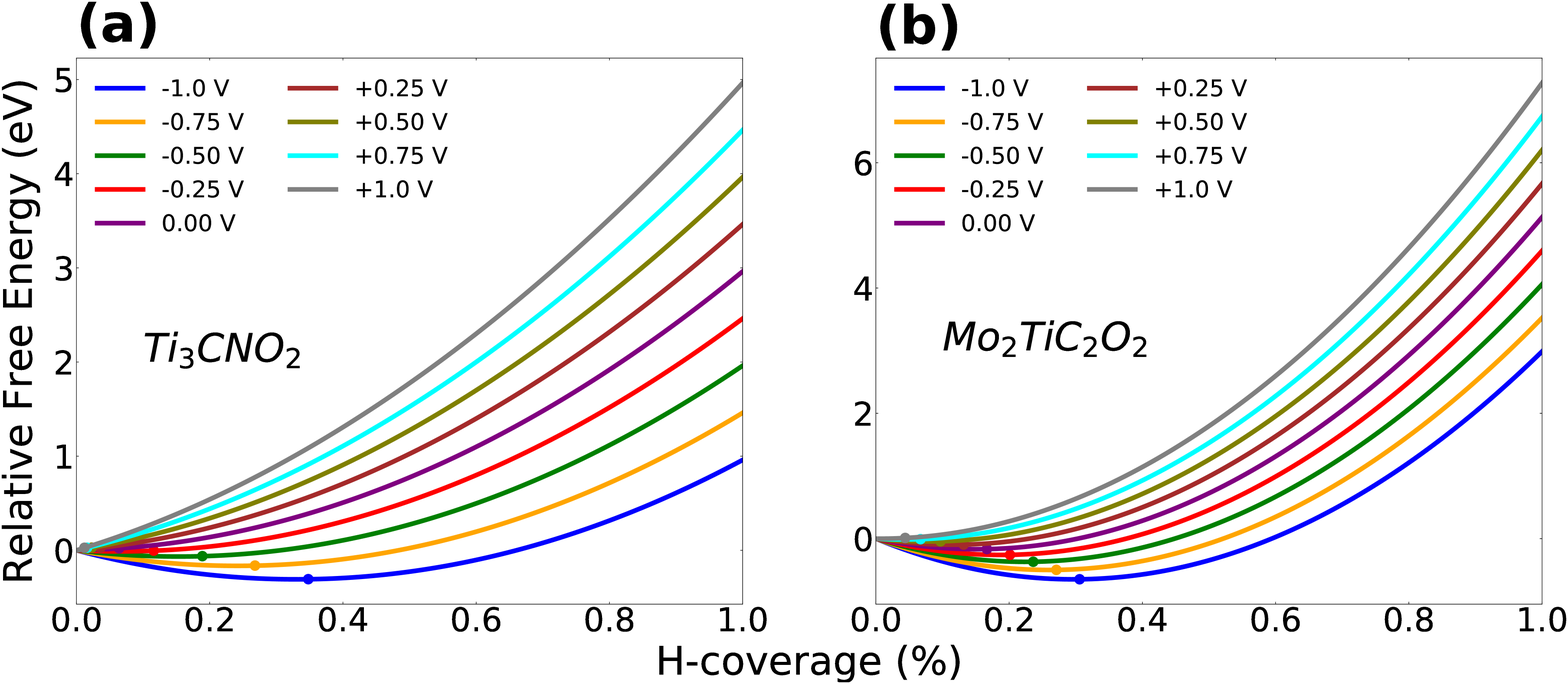}
    \vspace{0.1cm}  
    \includegraphics[width=0.95\textwidth,center]{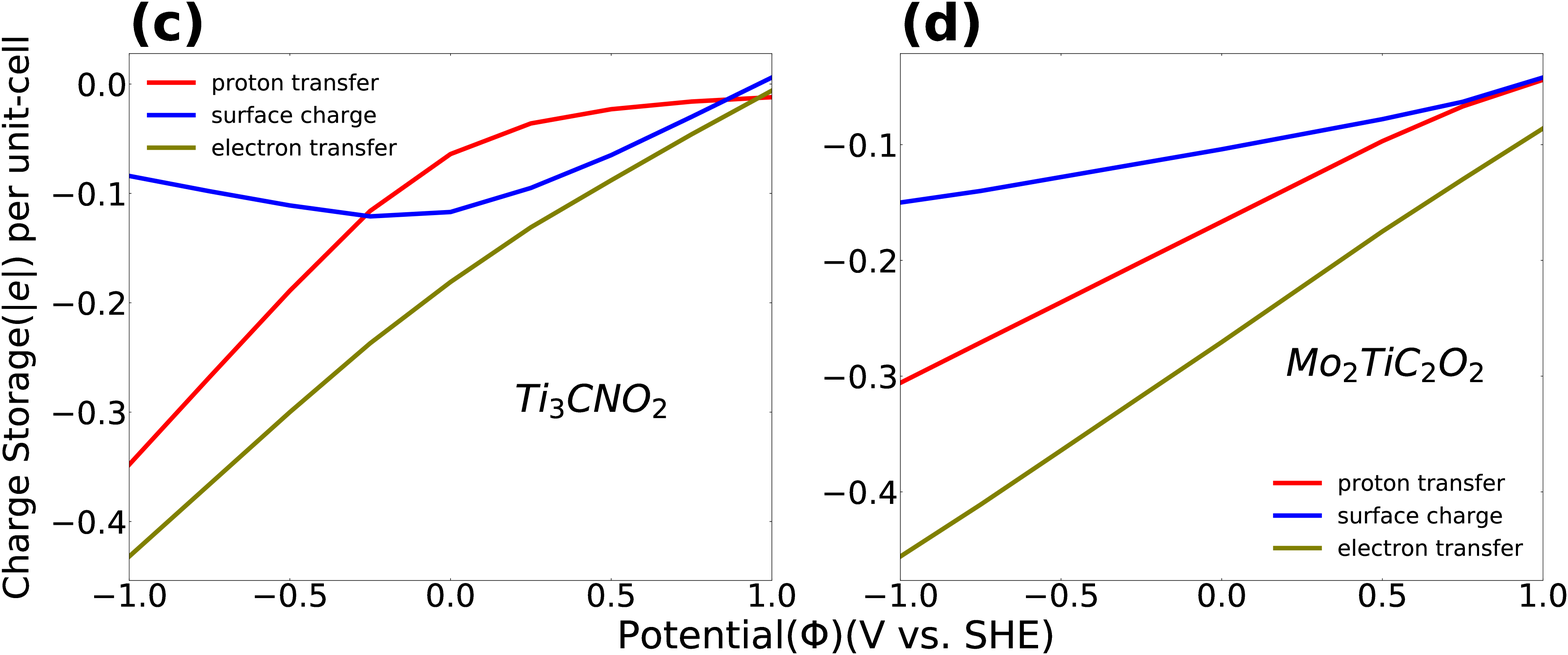} 
    \vspace{0.1cm}
    \includegraphics[width=0.95\textwidth,center]{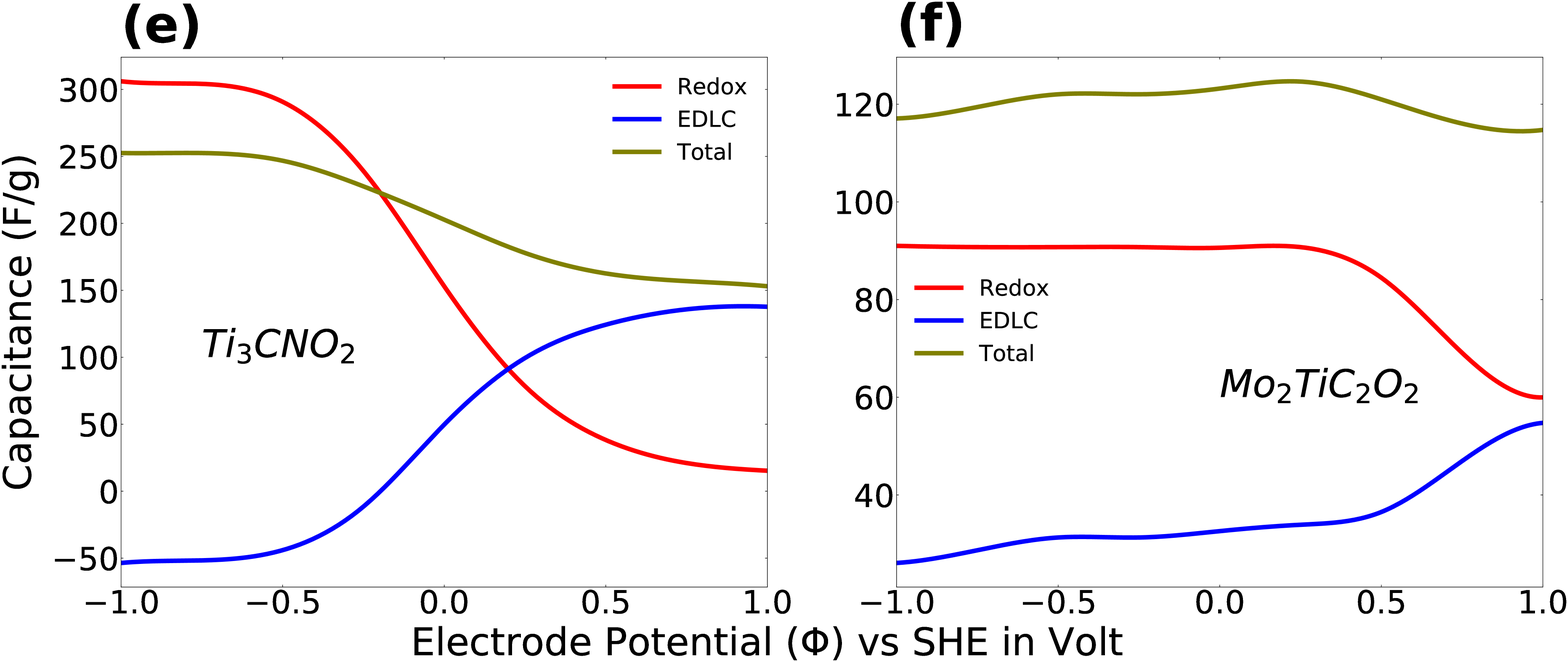}       
    \caption{Variations in the Relative Free Energy with H-coverage at different fixed potentials for $Ti_3CNO_2$(a) and $Mo_2TiC_2O_2$(b).The charge storage and electrical capacitances as a function of electrode potential are shown for $Ti_3CNO_2$(c,e) and $Mo_2TiC_2O_2$(d,f)}
    \label{Fig:7}
\end{figure*}

In case of the pristine system, total electric capacitance is nearly constant at 250 F/g between -1 V and 0 V. It decreases linearly to 200 F/g for higher potentials. Both qualitative and quantitative agreements with the existing result \cite{zhan2018understanding} on the variation in the total capacitance with $\Phi$ are excellent. For the LS doped system, the total capacitance is near constant at a value of 250 F/g throughout the range of the voltage. This is because of the near constant contributions from both EDLC and pseudocapacitance which in turn is due to the linear variations in proton transfer number and the surface charge. Therefore, nitrogen doping at the lattice site does not lead to any improvement in the electrical capacitance of $Ti_3C_2O_2$. The doping at the functional site, on the other hand, improves the total capacitance. This is due to higher value of $C_{redox}$ and lower value of $C_{EDL}$ for this system. Since in these cases, EDLC acts against pseudocapacitance, a lower value of $C_{EDL}$ implies that the surface electrochemistry is overwhelmingly determined by the redox mechanism. In case of the pristine and LS doped $Ti_3C_2O_2$, the EDLC, both qualitatively and quantitatively, contribute substantially to the total electrical capacitance. In these two cases, though redox appears to be the dominating mechanism, the effect of EDL is recognizable. The dominance of redox reaction as found in our calculations, however, is in contrast with the predictions made from the experimental results \cite{Ndopedti3c2,Ndopedti3c2-1}.

A qualitative understanding of the larger value of redox capacitance for FS doped system as compared to the LS doped one can be understood from  Bader charge analysis \cite{bader}. With full H-coverage, the average Bader charge transfer from the electrode to the $H^+$ ion is 0.463$e$, 0.501$e$ and 0.455$e$ for pristine, FS doped, and LS doped systems, respectively. The largest charge transfer for FS doped system can explain the largest pseudocapacitance. The reason FS doped system has a larger charge transfer than the LS doped one is the following: from Table \ref{tab:1}, we find that the Ti-N bond length for FS doped system is 15\% less than that in the LS doped system. In fact Ti-N and Ti-O bond lengths are comparable for FS doped systems as doped nitrogen occupies the functional sites. Due to significantly smaller Ti-N bond length, the charge sharing in Ti-N bond is greater in FS doped system than that in LS doped one. This, in turn, is responsible for larger charge transfer to the $H^+$ ions as can be seen quantitatively from calculation of Bader charges on each atom.
\subsubsection{Substituted $Ti_3C_2O_2$}
The electrochemical behaviours of the substituted systems Ti$_3$CNO$_2$ and Mo$_2$TiC$_2$O$_2$ throw up quite an interesting picture.  Figure ~\ref{Fig:7}(a)-(b) show the $G(x,\phi)$ as a function of H-coverage  for $Ti_3CNO_2$  and $Mo_2TiC_2O_2$, respectively. Unlike the doped systems, $Ti_3CNO_2$ and $Mo_2TiC_2O_2$ have significantly low coverages, the highest $x_{avg}$ being 0.35 and 0.31 respectively at -1 V. Behaviour of charge storages (Figure \ref{Fig:7} (c)-(d)) too, are very different from the doped systems.The magnitude of the charges are less for substituted systems in comparison to the pristine and doped systems.However, the surface charge and protons do not act against each other. Lower amount of charge storage (in comparison with doped systems) and the negative sign of the surface charge (unlike the doped systems) can be correlated with the substantially low H-coverage. The qualitative and quantitative variations of various components in charge storage are reflected in the behaviour of EDLC and pseudocapacitances. The non-linear variations of the proton and surface charges in Ti$_3$CNO$_2$ determine the variations of the capacitances. While pseudocapacitance was the dominant contributor for negative voltages, EDLC starts to play a significant role with increasing voltage, finally becoming the dominant one for positive voltages. Thus, in nitrogen substituted Ti$_3$C$_2$O$_2$, the mechanism of charge storage is dependent on the voltage window. Our quantitative calculations are in agreement with the quantitative analysis that Ti$_3$CNO$_2$ cathode has surface redox reaction as the dominating mechanism \cite{ti3cn}.
Behaviour of the components in electrical capacitance of Mo$_2$TiC$_2$O$_2$ is found to be different from all other systems. In here, the pseudocapacitance and EDLC do not oppose each other but collaborate. While pseudocapacitance turns out to be dominant, and both components remain near constant for most of the voltage window, beyond +0.5 V EDLC starts to assume more significant role. However, the absolute values are the least among the systems considered. This can be understood from the magnitudes of the charge storage which in turn is connected to the lowest H-coverage and the fact that  Mo being 60\% more electronegative than Ti, is unable to transfer as much charge as Ti in outer layers of the MXene.  
\subsection{Total Capacitance and comparison with experiments}
\begin{figure}[h]
    \includegraphics[width=0.55\textwidth,center]{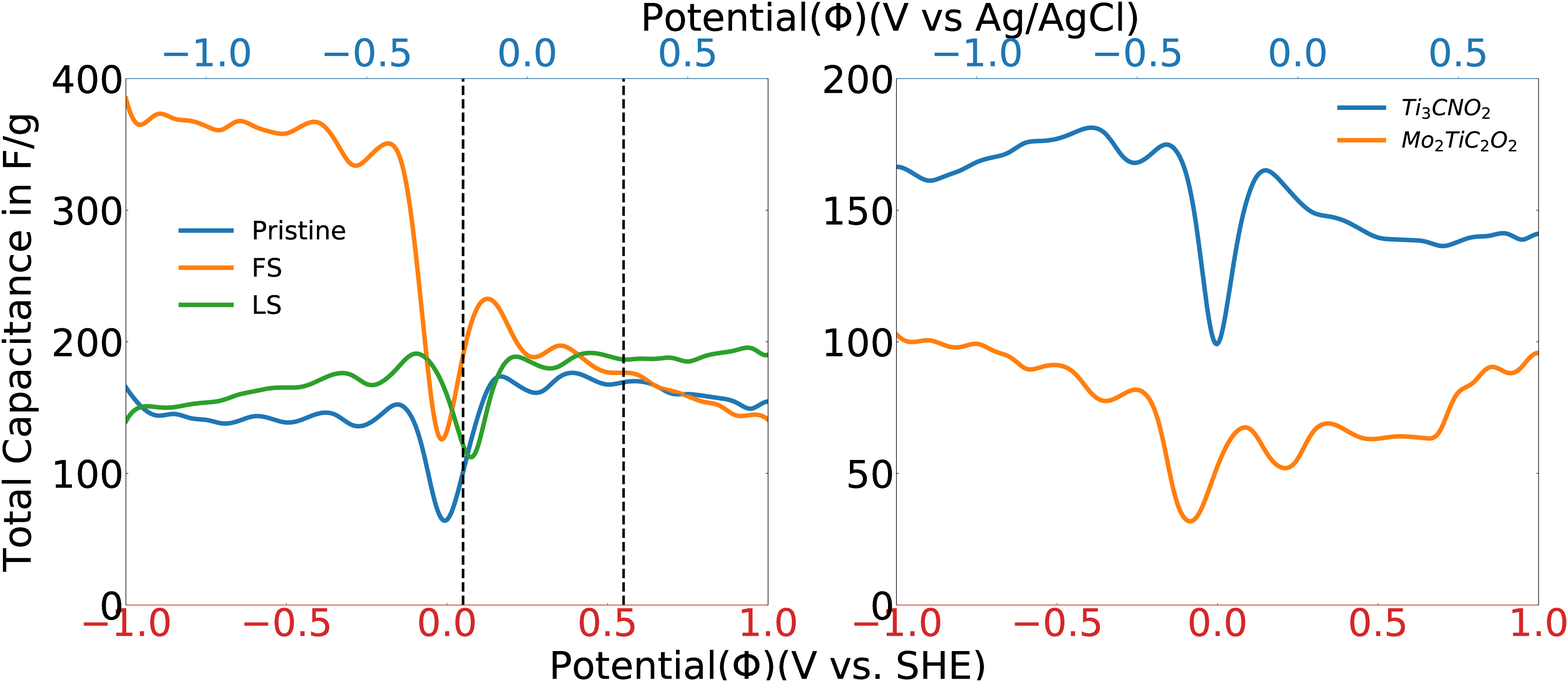}
    \caption{Total Capacitance as a function of  electrode potential for nitrogen-doped (left) and substituted (right) $Ti_3C_2O_2$. The vertical dotted lines in the left panel indicate the potential window used in the experiment on nitrogen doped $Ti_3C_2O_2$ \cite{Ndopedti3c2}.}
    \label{Fig:8}
\end{figure}

In Figure \ref{Fig:8} we show the total capacitance (calculated by Equation (\ref{Eqn:1})of pristine and doped Ti$_3$C$_2$O$_2$(left) ,$Ti_3CNO_2$ and $Mo_2TiC_2O_2$(right). The variations in the capacitances are plotted with respect to  SHE and Ag/AgCl electrode potentials. We find that the capacitance profiles for pristine and doped systems follow that of their quantum capacitances. This was to be expected as the orders of magnitude of $C_{E}$ and $C_{Q}$ are same for all systems except FS doped one at negative electrode potentials. As a consequence, we obtain quite high capacitance for FS doped system in the negative potential range. For the positive part of the SHE potential window the capacitances of pristine, FS doped and LS doped systems are comparable. An one to one  comparison with the experimental results is somewhat difficult because of various factors. The capacitance crucially depends on the sample preparation method, the exact composition, the functional groups and the electrolyte. Accordingly, our results are contradictory to the ones obtained in Reference ~\onlinecite{Ndopedti3c2-1} where the electrolyte is basic. The capacitances obtained in that work are significantly lower than our results. Also, they obtained the highest value of capacitance for LS doped system and identified EDL as the primary mechanism which is aided by the surface redox reactions due to the surface functional groups. On the other hand, some correspondences can be made with the experimental results of Reference ~\onlinecite{Ndopedti3c2} as the electrolyte used was 1 M H$_2$SO$_4$, exactly the one used in the model implemented in our calculations. Although the maximum capacitance reported by them is only 32 F/g for the pristine system as opposed to 176 F/g in our calculations, there is good agreement for the maximum capacitance of nitrogen doped system. The reported value of 192 F/g in the Ag/AgCl electrode potential window of -0.35-0.35 eV agrees well with the calculated values for FS (230 F/g ) and LS (192 F/g) doped systems in the same window. However, there are certain qualitative differences between the results from the experiment and our calculations. From the high resolution XPS study it was suggested that the doped nitrogens occupy the carbon sites. Therefore the capacitance reported in their work is that of a LS doped system. It is yet to be observed whether further gain in capacitance can be obtained in a FS doped system in agreement with our results. The experiment also suggested that this gain in capacitance is due to larger inter-layer distance due to doping leading to an increase in EDLC and therefore the primary mechanism in the doped system is EDL. Our calculations could not confirm it. The EDL part of the capacitance, in our calculations, rather went the opposite way. However it is worth mentioning that our calculations are on a monolayer implying infinite inter-layer distances. Moreover no other functional group except -O was used in our model while the experimental sample had  -F functional groups in significant proportions. The presence of -F groups that are less redox active than -O in the experimental sample may explain the discrepancy between the experiment and theory in identifying the primary mechanism for charge storage. The total capacitance of the substituted systems are less than the doped systems. While the variation in the total capacitance of Mo$_2$TiC$_2$O$_2$ follows that of its quantum capacitance, the situation is different for Ti$_{3}$CNO$_{2}$. Though it had larger quantum capacitance for higher voltages, the drop in the electrical capacitance with increasing voltage brings the maximum of total capacitance in the negative part of SHE voltage window. These results show that the quantum capacitance is an important component having substantial influence on the quantitative estimation of capacitances of supercapacitors as observed earlier \cite{das2022}.   
\section{Conclusions}
Using an implicit solvation model in conjunction with DFT we have modelled the electrolyte electrode interaction at the electrode surface for computation of the electrochemical parameters of doped and substituted oxygen functionalised Ti$_3$C$_2$ MXene. This model addresses the surface EDL and redox effects in H$_2$SO$_{4}$ electrolyte solution. We have computed the contributions to stored charges and the capacitances due to both effects when nitrogen is doped at different sites of Ti$_{3}$C$_{2}$O$_{2}$ as well as in substituted systems Ti$_{3}$CNO$_{2}$ (nitrogen substitution at Carbon site) and Mo$_{2}$TiC$_{2}$O$_{2}$ (molybdenum substitution at outer Ti layers). We find that upon inclusion of the quantum capacitance, the maximum gain in the capacitance is obtained for nitrogen doped at functional sites of Ti$_3$C$_2$O$_2$ at the negative SHE potentials. This two-fold gain, in comparison to the pristine compound is due to larger storage of charge dominated by the redox activity at the surface. The presence of nitrogen dopants at the surface along with the largest coverage of H$^{+}$ ions from the electrolyte are responsible for this. Our quantitative results suggest that for the pristine and doped systems the surface redox activity is primary responsible for the electrochemical parameters although EDL mechanism competes with it. The two mechanisms act against each other and influence the overall charge storage and electrical part of the capacitance. Quantitatively our results are in excellent agreement with the experiment \cite{Ndopedti3c2}. However, the qualitative explanation of increased capacitance value differs with ours. With increase in nitrogen content such that it substitutes 50\% carbon from the lattice site, the capacitance value degrades. It further degrades upon substituting the surface Ti atoms with more electronegative Mo.These are yet to be verified through electrochemical measurements although both compounds have been synthesised \cite{ti3cn-1,Mo2TiC2}. In cases of substitution, though EDL and surface redox phenomena co-operate each other, poorer charge transfer hinders growth of capacitances. In case of nitrogen substitution we found evolution of EDL mechanism to become the dominant one upon increase in voltage. The reverse trend in comparison to doped systems can be correlated with remarkably low ion coverage from the electrolyte and subdued charge transfer in the surface due to presence of more electronegative substituents. Our results suggest that nitrogen doping is a better strategy to get better electrochemical performances in Ti$_3$C$_{2}$O$_{2}$ MXene electrodes.We also suggest that much better values of capacitances in nitrogen doped Ti$_{3}$C$_{2}$O$_{2}$ can be obtained if samples are prepared with doping at the functional site and electrochemical measurements are done in a larger negative voltage window with Ag/AgCl as reference electrodes. 

\section{Acknowledgement}
The authors gratefully acknowledge the Department of Science and Technology, India, for the computational facilities under Grant No. SR/FST/P-II/020/2009 and IIT Guwahati for the PARAM supercomputing facility. 



%

\end{document}